\documentclass[a4paper,11pt]{article}
\usepackage{subcaption}
\usepackage{pos}
\usepackage{lipsum} 
\usepackage{wrapfig}
\usepackage{lipsum} 
\usepackage[version=4]{mhchem}
\usepackage{lineno}
\usepackage{cleveref}
\usepackage{siunitx}
\title{Very Late Afterpulses and Search for the Neutron Echo in IceCube}

\author{The IceCube Collaboration \\{\normalsize \normalfont(a complete list of authors can be found at the end of the proceedings)}\\}

\emailAdd{kdutta@icecube.wisc.edu}
\emailAdd{sboeser@icecube.wisc.edu}
\emailAdd{mrongen@icecube.wisc.edu}

\abstract{

While high-energy astrophysical neutrinos are well-established, their flavor composition remains relatively unconstrained. In IceCube, long muon tracks from $\nu_{\mu}$-CC interactions are easily identified but the detector geometry does not allow sufficient granular resolution to distinguish the cascade-type events. The Neutron Echo - a delayed light signal in the detector from neutron capture and de-excitation - can probe the shower’s hadron content and thus the underlying interaction. A significant background arises from the late PMT afterpulses, which are temporally coincident with the physics signal. The traditional IceCube DAQ has a limited readout window with significant deadtime between triggers, which is insufficient to capture the late pulses. A recently developed deadtime-free DAQ mode, with an extended readout window, enables their detection. An observed excess in the delayed time spectrum over the background would be compatible with the Neutron Echo hypothesis. \\
In this contribution, we summarize the physics scope of delayed signals, discuss the timing spectrum of the signal and PMT background, highlight the capabilities of the new DAQ system for recording late pulses, and emphasize the potential of IceCube for particle identification through delayed signals.

\vspace{4mm}

{\bfseries Corresponding authors:}
Kaustav Dutta$^{1*}$, 
Sebastian Böser$^{1}$, 
Martin Rongen$^{2}$\\
{$^{1}$ \itshape Johannes Gutenberg Universität, Saarstraße 21, 55122 Mainz, Germany}\\
{$^{2}$ \itshape Erlangen Centre for Astroparticle Physics, Friedrich-Alexander-Universität Erlangen-Nürnberg}\\[4mm]
$^*$ Presenter
}

\ConferenceLogo{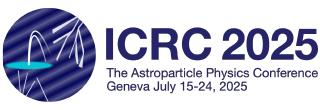}

\FullConference{39th International Cosmic Ray Conference (ICRC2025)\\
 15–24 July 2025\\
Geneva, Switzerland\\}

\let\oldthebibliography\thebibliography
\let\endoldthebibliography\endthebibliography
\renewenvironment{thebibliography}[1]{%
  \oldthebibliography{#1}%
  \setlength{\itemsep}{0pt}
  \setlength{\parskip}{0pt}
}{%
  \endoldthebibliography}%

\begin{document}

\maketitle

\section{Physics Scope of Delayed Light Emission Processes}\label{sec1}
\indent The first direct detection of $\bar{\nu}_e$ via the inverse beta decay (IBD) reaction $\bar{\nu}_e + p \rightarrow e^+ + n$ was achieved by Cowan and Reines in 1956 \cite{Cowan:1956} near the Savannah River Plant in South Carolina. The positron quickly produced an annihilation signal, while the thermalized neutrons were captured by cadmium nuclei, emitting delayed photons in the process. The coincidence of the prompt positron signal followed by the delayed neutron capture signal enabled a clear identification of $\bar{\nu}_e$ and substantial background suppression. Since then, delayed neutron captures have become a standard technique for isolating rare events in high-background environments \cite{Watanabe2009NeutronTagging,DayaBay:2012fng,DoubleChooz:2011kae}. Super-Kamiokande, for example, employed neutron tagging on hydrogen to search for supernova relic neutrinos (SRN) and reported 13 IBD candidate events below 30 MeV \cite{Super-Kamiokande:2015qek}.\\
\indent At higher energies, neutron capture signals can help probe astrophysical neutrino flavor composition, hinting at production mechanisms at the source. Under standard oscillations, an initial flavor ratio of $\left( \frac{1}{3} : \frac{2}{3} : 0 \right)$ at the source evolves to approximately $\left( \frac{1}{3} : \frac{1}{3} : \frac{1}{3} \right)$ at Earth due to vacuum mixing \cite{IceCube_flavor_range}. Since the expected flavor composition at Earth is tightly constrained \cite{IceCube_flavor_range}, any deviations may indicate alternative production mechanisms or new physics. However, distinguishing
$\nu_\tau$ from $\nu_e$ events remains challenging in neutrino telescopes due to insufficient spatial resolution to distinguish between the shower signatures\cite{IceCube_flavor_range,Seven_taus}.
Ref.~\cite{Li:2016gep} identifies delayed neutron capture signals as a promising technique for neutrino flavor discrimination. The method exploits the lower neutron yield in $\nu_e$-induced showers due to a much smaller hadronic component than the $\nu_\tau$ showers. A 9-fold improvement in flavor separation over conventional techniques has been reported.\\
\indent Identifying neutrino interaction channels with delayed signals also helps in understanding the energy dependence of the astrophysical neutrino flux which follows a power-law behavior characterized by the spectral index $\gamma$, i.e., $\phi(E_\nu) \propto E_\nu^{-\gamma}$ \cite{IceCube:2013cdw}. At a given reconstructed event energy, the contribution of neutral current (NC) interactions from all flavors depends on both the assumed $\gamma$ and reconstructed $E_\nu$ due to the steeply falling neutrino flux and the inelasticity distribution which peaks at low values (average inelasticity $\approx$ 0.3 for a combined flux of $\nu + \bar{\nu}$ at 100~TeV \cite{IceCube_100TeV_statement}). At the IceCube benchmark spectral index $\gamma \approx 2.6$ \cite{spectral_index}, NC interactions are expected to contribute about 7\% of events at 100~TeV. An excess of high-energy hadronic showers could serve as a potential signature of new physics scenarios, such as the presence of boosted dark matter \cite{boosted_dark_matter}. In such models, a heavy dark matter particle may decay into a lighter secondary particle that scatters off a nucleon within the IceCube detector, producing a purely hadronic shower. The ability to distinguish shower types using delayed photon signatures offers a promising opportunity to validate such events.

\section{Neutron Echo in IceCube}\label{sec2}
\begin{figure}[htbp]
\vspace{-15pt} 
    \centering
    \begin{minipage}[t]{0.48\textwidth}
        \centering
        \includegraphics[width=\linewidth]{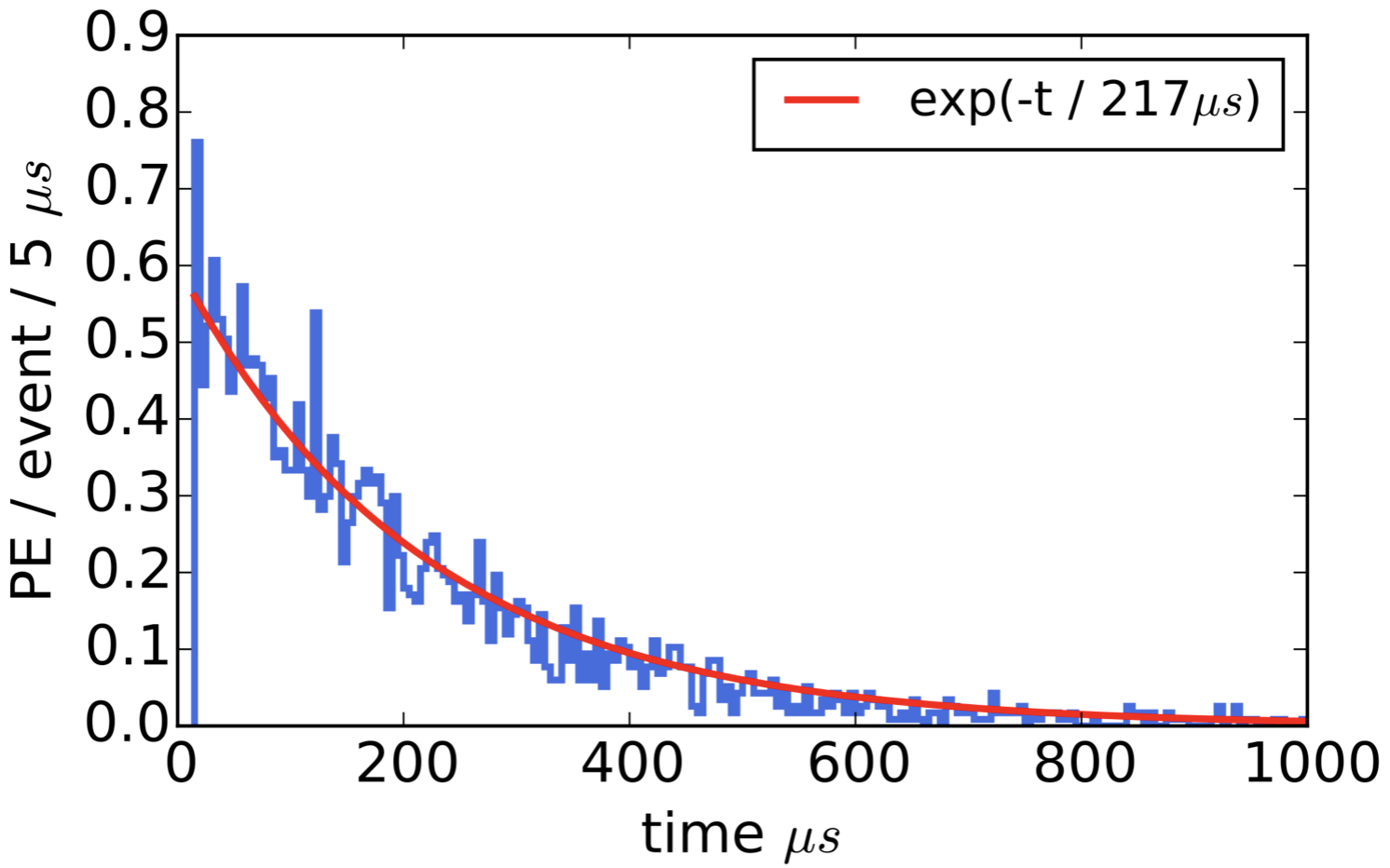}
    \end{minipage}%
    \hfill
    \begin{minipage}[t]{0.48\textwidth}
        \centering
        \includegraphics[width=\linewidth]{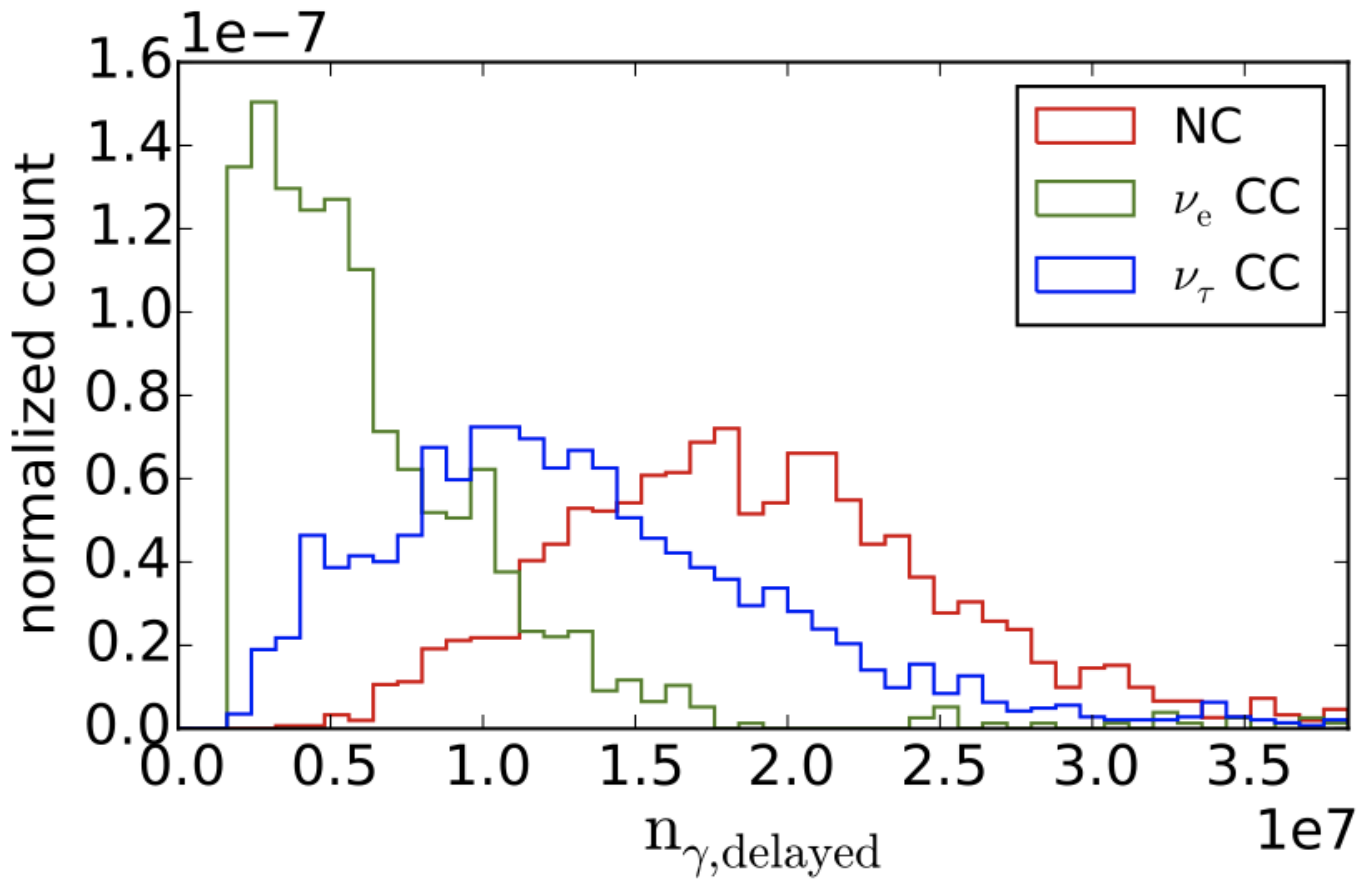}
    \end{minipage}
    \caption{Left: Timing spectrum of delayed photons incident on modules within 15–1000~$\mu$s from Geant4 simulations of IceCube NC events, before applying detector effects such as deadtime; Right: Delayed photon counts from Geant4 simulations for shower-like events between 100–150~TeV. Both taken from \cite{Steuer:PhDthesis}.}
    \label{fig:neutron_echo_combined}
\end{figure}

\indent Every neutrino interacting with ice initiates a hadronic shower through nuclear fragmentation. The free neutrons produced in hadronic interactions thermalize by multiple inelastic scatterings with the ice nuclei, which typically occurs within a few microseconds. Once thermalized, approximately 99\% of the neutrons are captured on hydrogen atoms, forming deuterium through the reaction $n + p \rightarrow d + \gamma$, where the emitted \SI{2.2}{\mega\electronvolt} photons produce Compton electrons. These Compton electrons in turn emit Cherenkov radiation, which is detected as the delayed neutron echo signal. The neutron capture lifetime in ice is approximately \SI{217}{\micro\second} \cite{Steuer:PhDthesis}, which defines the characteristic delay of the neutron echo signal (Fig.~\ref{fig:neutron_echo_combined} [left]) relative to the prompt Cherenkov emission, which are detected within the first hundred nanoseconds. \\

\indent The relative intensity of the neutron echo scales with the neutron abundance, which is highest in hadronic showers from NC events. Next are $\nu_\tau$ CC interactions where a hadronic shower occurs at the primary vertex and the tau decay vertex can be either hadronic or EM. The $\nu_e$ CC events generate a weaker delayed photon signal due to fewer shower neutrons produced in photonuclear interactions. \texttt{GEANT4} simulations of high-energy cascades in IceCube (Fig.~\ref{fig:neutron_echo_combined} [right]) demonstrate the potential of the neutron echo to distinguish between interaction channels. The delayed photon distribution exhibits a skewed Gaussian shape for CC events which follow the inelasticity distribution, and a symmetric Gaussian profile for the NC events. A clear separation of shower types is evident.
\section{Detector response}\label{sec3}
\indent The fundamental detection unit in IceCube is the Digital Optical Module (DOM), which captures incident photons via a photomultiplier tube (PMT). The photons are incident at the photocathode, which emits photoelectrons that are amplified through dynode stages before reaching the anode plate. A discriminator starts the data acquisition (DAQ), also referred to as a "DOM launch", when the signal amplitude exceeds 0.25 photoelectrons (PE) \cite{IceCube_detector}. To capture these signals, each module is equipped with three onboard digitizers. The two Analog Transient Waveform Digitizer (ATWD) chips have a fast sampling rate of $\approx$ 330\,MHz and are each equipped with 3 gain channels to completely capture the dynamic range of the PMT output. The ATWDs record waveforms over a \SI{427}{\nano\second} window offering sufficient resolution for prompt Cherenkov light detected within tens of meters of the interaction vertex. To detect longer-duration, lower-amplitude signals from more distant DOMs, a \SI{40}{\mega\hertz} fast Analog-to-Digital Converter (fADC) continuously samples analog signals in 6.4 $\text{\textmu s}$ readout windows. Since the neutron echo is expected to be a faint signal, data from saturated DOMs are important for its detection. The digitization operation in such scenarios leads to deadtime effects as discussed further.\\

\newpage

\subsection{Deadtime effects in digitizers}

\begin{wrapfigure}[15]{r}{0.48\textwidth}
\vspace{-5mm}
    \centering
    \includegraphics[width=\linewidth]{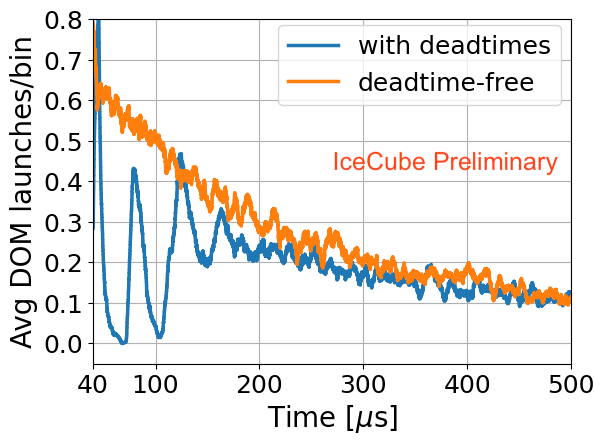}
    \caption{DOM launch rates averaged over 500 FBLED flashes at a brightness setting b25. Standard DAQ with ATWD-induced deadtime is shown in blue; deadtime-free mode in orange.}
    \label{fig:dom_launches_deadtime_free}
\end{wrapfigure}

\indent Data acquisition by the fADC and ATWDs is synchronized. The fADC digitizes continuously over a \SI{6.4}{\micro\second} window, while the ATWD captures and sequentially digitizes each gain channel requiring \SI{\sim30}{\micro\second} per channel \cite{IceCube_detector}. During digitization, the ATWD chip is unavailable for further data acquisition. After the first fADC readout window, a brief reset occurs before acquisition resumes with the second ATWD chip. For bright events, digitization of all three gain channels in the first ATWD takes up to $3 \times 30\,\text{\textmu s} = 90\,\text{\textmu s}$. The second ATWD typically sees a weaker signal requiring only one channel and \SI{\sim30}{\micro\second}. This introduces a deadtime of \SI{\sim30}{\micro\second} - \SI{6.4}{\micro\second} = \SI{23.6}{\micro\second} during which both ATWDs are unavailable. Due to synchronization, this deadtime is also imprinted in the fADC launch distribution (Fig.~\ref{fig:dom_launches_deadtime_free}). \\
\indent The deadtime effects depend on the number of used amplification channels which correlates with the distance of the receiving DOM from the source. Previous studies \cite{Steuer:PhDthesis,meinusch_deadtimes} have shown that the deadtime effects are complicated and difficult to model analytically. These deadtimes distort the waveform shape and prevent a detailed study of late PMT effects which are backgrounds to the Neutron Echo signal. To address this, the DAQ firmware was modified to include a custom-made deadtime-free fADC-only mode. This mode, developed specifically for the late pulse study, is not a part of the standard IceCube physics data-taking. In this configuration, the ATWD readouts are skipped and only the continuously digitized fADC waveforms are recorded. This removes the 30 $\text{\textmu s}$ deadtime per ATWD channel and allows for full waveform recovery, as shown in orange in Fig.~\ref{fig:dom_launches_deadtime_free}. The local coincidence requirements between neighboring DOMs are also disabled allowing analysis of the complete timing distribution from both Cherenkov and DOM-induced photons for each in-situ DOM. The background effects from the latter are now further discussed.
\subsection{Delayed pulses from PMT \& DOM glass housing}
\label{DOM_PMT_bkg}

\indent An electric field inside the PMT is established by a voltage divider chain \cite{IceCube_PMT_paper}, which directs photoelectrons toward a series of dynodes for signal amplification. During standard amplification, the prompt Cherenkov signal is recorded in the PMT waveform within tens of nanoseconds after the initial trigger. Occasionally, the photoelectrons backscatter toward the photocathode from the first dynode, are re-accelerated by the electric field, and strike the dynode again after a delay to initiate the amplification cascade. The delay is equal to the transit time between the photocathode and dynode, typically ranging between \SI{10}{\nano\second} and \SI{80}{\nano\second} \cite{IceCube_detector}. Residual gas molecules inside the PMT may also become ionised by the accelerated photoelectrons during the amplification process \cite{IceCube_PMT_paper}. The positive ions drift back toward the photocathode and release secondary electrons upon impact. These produce delayed signals known as afterpulses, occurring several microseconds after the prompt pulse. The delay depends on the mass of the ion which determines its drift time. In IceCube PMTs, prominent afterpulses are observed at 2~$\text{\textmu s}$ and 8~$\text{\textmu s}$, attributed to \ce{He+} and \ce{Cs+}, respectively \cite{IceCube_PMT_paper}. 
\begin{wrapfigure}[14]{l}{0.48\textwidth}
\vspace{-10pt} 
    \centering
    \includegraphics[width=\linewidth]{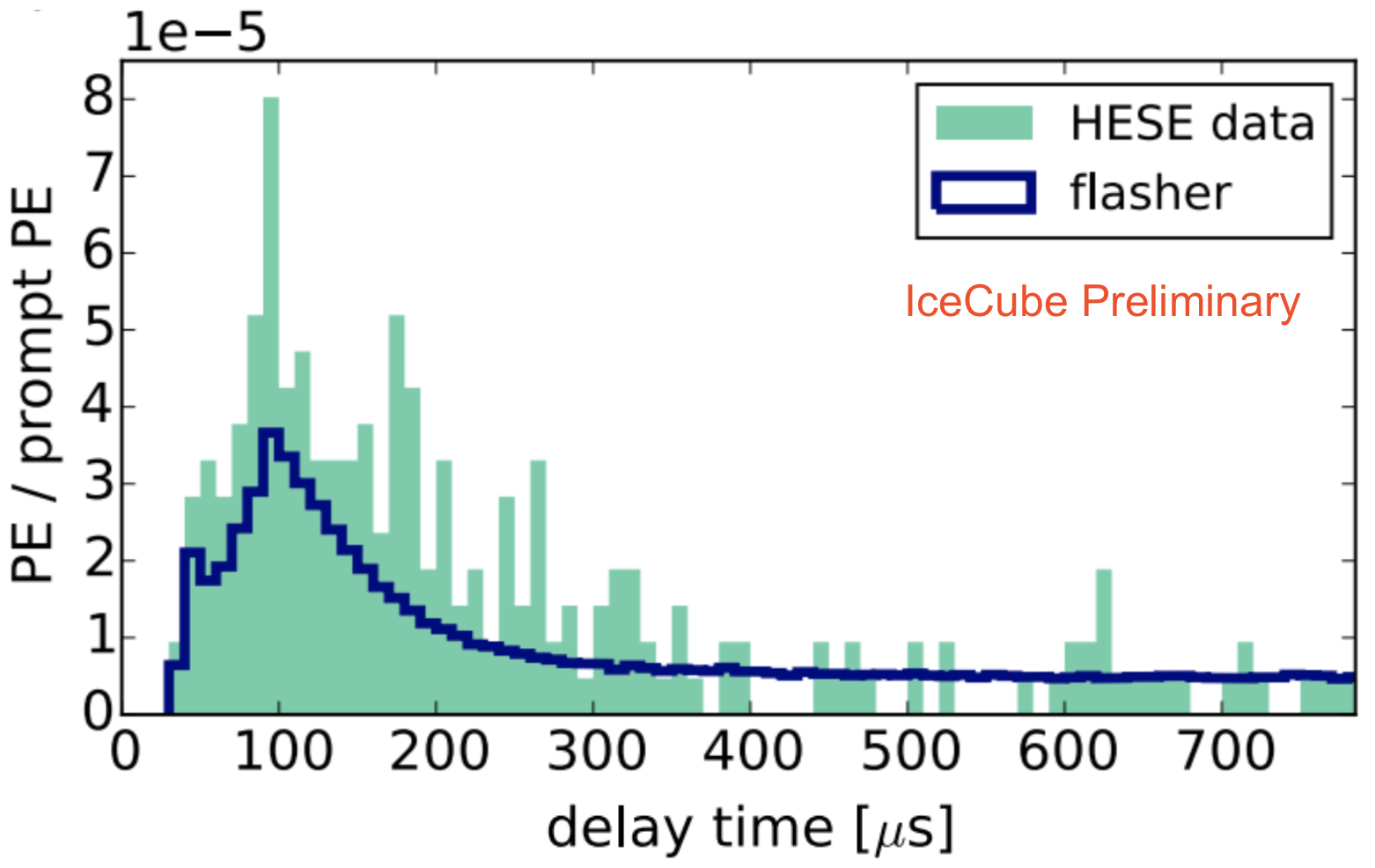}
    \caption{Stacked time distribution of delayed photons from 1250 high energy starting events (HESE) \cite{HESE} events recorded between 2015 and 2019, alongside flasher run data from two in-situ DOMs.}
    \label{fig:Neutron_echo_background}
\end{wrapfigure}
\indent Independent measurements in 8-inch Hamamatsu and Thorn PMTs also found unexpected pulses roughly 100~$\text{\textmu s}$ after the prompt pulse \cite{Thorn_afterpulses}. The excess signal was attributed to PMT-intrinsic effects since they were measured in bare PMTs, although the exact cause remains unclear. An ionic explanation suggests that these pulses originate from a heavy molecule about 150 times the mass of \ce{Cs+}, possibly bialkali metals embedded in the photocathode \cite{bialkali} or glass envelope , or particulate contamination inside the PMT. Alternatively, delayed signals could arise from electron trapping at impurity sites or defects within the photocathode material \cite{electron_trapping}. The pulse intensity was observed to scale with the prompt signal strength and was predominantly composed of single photoelectrons (SPEs). Following this discovery, dedicated flasher LED runs were conducted with two IceCube DOMs which detected similar artifacts around \SI{100}{\micro\second}. The timing distribution of these late pulses closely matched the Neutron Echo spectrum (Fig.~\ref{fig:Neutron_echo_background}) making it unclear whether any excess signal above the PMT background could be attributed to neutron-induced light. The unexpected background halted the Neutron Echo analysis and therefore has been a focus of this study.
\begin{wrapfigure}[18]{r}{0.48\textwidth}
\vspace{-10pt} 
    \centering
    \includegraphics[width=\linewidth]{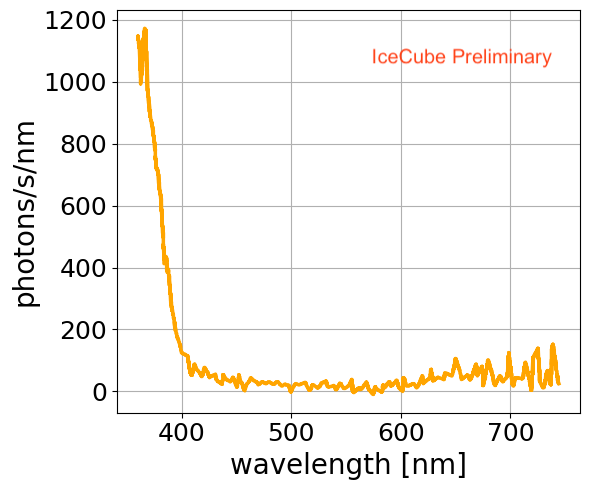}
\caption{Emission spectrum measured through IceCube Benthos glass, corrected for detector response and glass transmission. The \SI{265}{\nano\meter} excitation light is fully absorbed by the glass, and the detected signal arises from delayed re-emission within the glass.}
    \label{fig:benthos_luminescence}
\end{wrapfigure}

\indent Certain other processes have since been discovered which have similar lifetimes as the PMT-induced background described above. One example is photoluminescence from the borosilicate glass housing the PMT. The intrinsic defects and impurity centers within the glass, such as transition metal ions, absorb photons and become excited to metastable states. These centers then relax by emitting photons over prolonged timescales extending to hundreds of microseconds \cite{UV_luminescence_fit}, on the same order as the neutron echo lifetime. In a previous study of this effect, a 265\,nm LED equipped with a bandpass filter, to suppress any secondary luminescence from the LED itself, was directed onto an IceCube glass sphere. The 265\,nm light was completely absorbed by the glass, ensuring that only re-emitted light from the glass reached a spectrometer. The results indicated that luminescence is predominant in the ultraviolet (UV) region with a steep decline beyond 370 nm (Fig.~\ref{fig:benthos_luminescence}). This is particularly relevant to this study as the UV portion of the Cherenkov spectrum attenuates more rapidly with distance than longer wavelengths, leading to reduced luminescence-induced backgrounds at larger separations between source and module. The next section is dedicated to modeling the background components covered here.

\subsection{Analysis \& Modeling of delayed photon background}
\indent Recent runs within the IceCube detector involved flashing onboard LEDs and recording signals with the PMTs on the same module. The data acquisition was performed with both modes - fADCs configured for dead-time-free readout and ATWDs to capture the prompt signal intensity. A wide range of LED brightness settings was scanned, producing PMT signals up to several thousand PE to probe waveform characteristics across intensities. The 12 flashed modules were located at depths between 1300\,m and 2600\,m to investigate the influence of the PMT, glass housing, and surrounding ice on delayed pulse distributions. In addition to standard flashers emitting at 405 nm, specialized color DOM (cDOM) LEDs were flashed at four wavelengths — \SIlist{340;370;450;505}{\nano\meter} — to study the wavelength dependence of the background. Two high quantum efficiency DOMs were also flashed in the DeepCore \cite{IceCube_DC} region to examine possible photocathode effects.
\begin{figure}[htbp]
    \centering
\vspace{-5pt} 
    \begin{subfigure}[t]{0.48\textwidth}
        \centering
        \includegraphics[width=\textwidth]{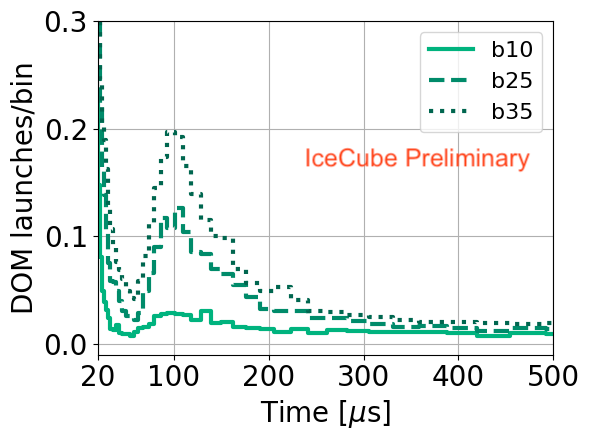}
        \label{fig:launches_brightness}
    \end{subfigure}
    \hfill
    \begin{subfigure}[t]{0.48\textwidth}
        \centering
        \includegraphics[width=\textwidth]{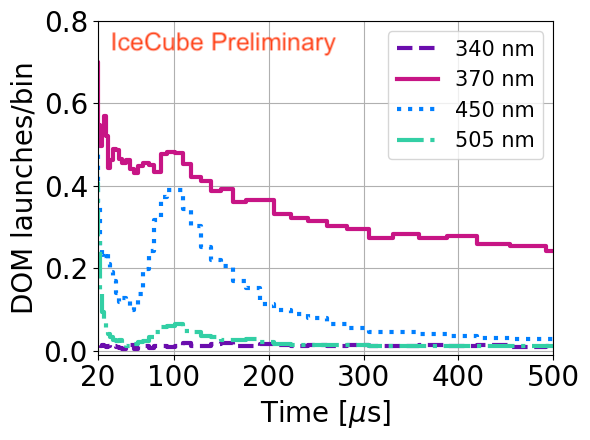}
        \label{fig:launches_wavelengths}
    \end{subfigure}
    \vspace{-1.5em}  
\caption{Left: DOM launch distributions with the deadtime-free DAQ mode from \SI{505}{\nano\meter} LED emission at brightness settings $\text{b}10$, b$25$, and b$35$, where the brightness intensity increases toward $\text{b}35$. Right: DOM launch distributions for LEDs at \SI{340}{\nano\meter}, \SI{370}{\nano\meter}, \SI{450}{\nano\meter}, and \SI{505}{\nano\meter}, all at $\text{b}15$. The \SI{340}{\nano\meter} and \SI{370}{\nano\meter} data have lower statistics due to the lower average photon yield of these LEDs.}
    \label{fig:launches_brightness}
\end{figure}
\begin{wrapfigure}[15]{l}{0.48\textwidth}
\vspace{-25pt} 
    \centering
    \includegraphics[width=\linewidth]{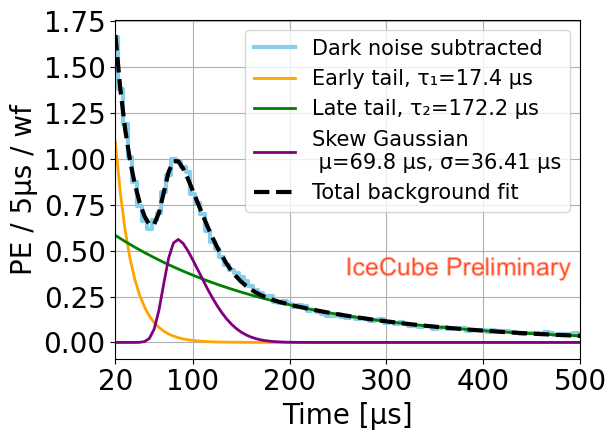}
    \caption{Average charge distribution from 5000 waveforms recorded for a module illuminated by a 405~nm flasher, with fit components and parameters. Each waveform corresponds to a single LED flash.}
    \label{fig:charge_distribution}
\end{wrapfigure}

\vspace{-8mm}

\begin{wrapfigure}[26]{r}{0.48\textwidth}
\vspace{-15pt} 
    \centering
    \includegraphics[width=\linewidth]{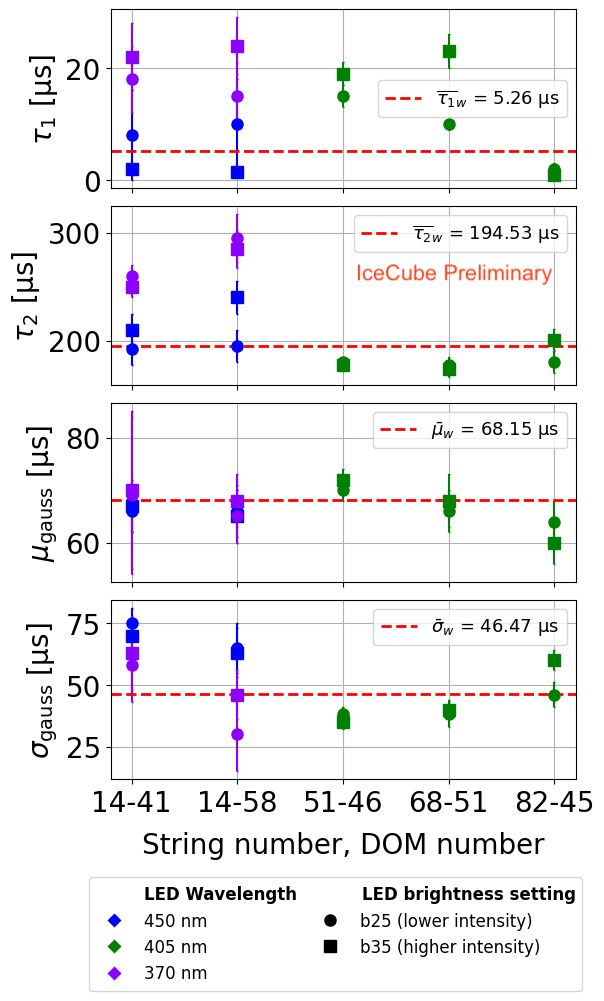}
\caption{Fitted parameters with uncertainties for five in-situ DOMs are shown versus signal intensity and LED wavelength. Weighted means $\bar{\tau}_{1w}$, $\bar{\tau}_{2w}$, $\hat{\mu}_{w}$, and $\hat{\sigma}_{w}$ are shown as red dashed lines.}
    \label{fig:fitting_parameters}
\end{wrapfigure}

\indent The DOM launch distributions with the deadtime-free mode show that delayed pulse intensity scales with signal strength (Fig.~\ref{fig:launches_brightness} [left]), supporting the measurements in \cite{Thorn_afterpulses}. A significant increase in launches around 100 µs is evident for the \SI{450}{\nano\meter} and \SI{505}{\nano\meter} settings (Fig.~\ref{fig:launches_brightness} [right]). This feature is less distinct at \SI{370}{\nano\meter} where the overall launch count remains consistently higher between \SI{20}{\micro\second}--\SI{1000}{\micro\second}, effectively drowning out the pronounced peak. These results suggest a simultaneous delayed light emission consistent with a luminescence-like process.\\
\indent For each brightness and wavelength setting in every module, all charge distributions (Fig.~\ref{fig:charge_distribution}) show an initial rapid decay (``early tail''), a PMT-induced Gaussian component, and a prolonged ``late tail'' extending up to 1~ms. In addition to glass luminescence discussed in Section \ref{DOM_PMT_bkg}, time-resolved studies of bare UV LEDs \cite{UV_luminescence_fit} also attribute the early and late tails to afterglow caused by defects in the semiconductor material. The intensity profile in the \SI{20}{\micro\second}–\SI{1000}{\micro\second} range is modeled using a double exponential. The Gaussian component was previously modeled as a skewed Gaussian in Thorn PMTs and in IceCube flasher data. Building on these observations, we model the full distribution as the sum of two exponentials and a skewed Gaussian function:
\vspace{-1.5mm}
\[
f(t) = A_1 e^{-t/\tau_1} + A_2 e^{-t/\tau_2} + A_g\, \text{SG}(t; \mu, \sigma, \alpha),
\]
where $\mathrm{SG}$ represents a skewed Gaussian function characterized by mean position $\mu$, width $\sigma$, and skewness $\alpha$. The parameters $A_1$, $\tau_1$ describe the early tail, $A_2$, $\tau_2$ the late tail, and $A_g$ scales the skewed Gaussian component. The fitted parameter trends for 5 DOMs - 2 cDOMs (string  14), 2 standard DOMs (strings 51, 68) and one DeepCore DOM (string 82) are presented in Fig.~\ref{fig:fitting_parameters} as functions of LED intensity and wavelength. All fits were performed over waveform intervals between \SI{30} and \SI{1000}{\micro\second}.

\indent  Although a fully coherent interpretation is challenging, several trends are still evident. For most DOMs, the late exponential tail lifetimes $\tau_2$ show a stronger dependence on wavelength than on illumination intensity. The background lifetimes from \SI{370}{\nano\meter} light are also generally longer than those observed at \SI{405}{\nano\meter} and \SI{450}{\nano\meter}. This behavior is consistent with expected luminescence effects, which peak at 370 nm and diminish sharply at longer wavelengths. Within the reported uncertainties, $\tau_1$ remains an order of magnitude shorter than $\tau_2$ at \SI{370}{\nano\meter}, consistent with the measurements of Ref.~\cite{UV_luminescence_fit}. The brightness- and wavelength-averaged $\tau_2$ value of \SI{194}{\micro\second} approaches the neutron echo lifetime ($\sim$ \SI{217}{\micro\second}), indicating potential challenges in discriminating signal from background templates. The Gaussian mean $\mu_\text{gauss}$ is consistently near \SI{68}{\micro\second} and largely independent of brightness and wavelength, suggesting a PMT-related origin. Due to fitting a skewed Gaussian, this mean is lower than the peak value which appears near \SI{100}{\micro\second}. No significant variation was observed in the Gaussian width, $\sigma_\text{gauss}$, which fluctuates around \SI{46}{\micro\second}.
\section{Summary \& Outlook}\label{sec4}
\indent Neutron-induced signals potentially provide an improved flavor and channel classification of cascade showers over conventional methods. Delayed light from the glass and PMTs within IceCube modules emerge as significant backgrounds to the Neutron Echo signal. The intensity of this delayed background light spectrum between \SI{30} and \SI{100}{\micro\second} was observed to scale with the prompt signal intensity. This distribution is modeled using a combination of a double-exponential and a skewed Gaussian function. The exponential component is consistent with luminescence originating from the LED, glass, or both. The Gaussian component remains temporally stable, centered around $\sim$ \SI{68}{\micro\second}, and is associated with a PMT-induced effect, showing no dependence on intensity or wavelength. The long exponential lifetime closely matches the expected Neutron Echo lifetime.\\
\indent Ongoing work aims to quantify the rate of delayed pulses relative to prompt signals which requires a reliable calibration of the main signal, and is currently hindered by PMT waveform saturation. The subsequent analysis of background-subtracted data will enable evaluation of flavor-dependent sensitivities associated with the Neutron Echo signal.

\begingroup
\small            
\setlength{\baselineskip}{0.9\baselineskip}  
\bibliographystyle{ICRC}
\bibliography{references}
\endgroup

%

\clearpage

\section*{Full Author List: IceCube Collaboration}

\scriptsize
\noindent
R. Abbasi$^{16}$,
M. Ackermann$^{63}$,
J. Adams$^{17}$,
S. K. Agarwalla$^{39,\: {\rm a}}$,
J. A. Aguilar$^{10}$,
M. Ahlers$^{21}$,
J.M. Alameddine$^{22}$,
S. Ali$^{35}$,
N. M. Amin$^{43}$,
K. Andeen$^{41}$
% ...
\\
\\
% ...
$^{10}$ Universit{\'e} Libre de Bruxelles, Science Faculty CP230, B-1050 Brussels, Belgium \\
% ...
$^{16}$ Department of Physics, Loyola University Chicago, Chicago, IL 60660, USA \\
$^{17}$ Dept. of Physics and Astronomy, University of Canterbury, Private Bag 4800, Christchurch, New Zealand \\
% ...
$^{21}$ Niels Bohr Institute, University of Copenhagen, DK-2100 Copenhagen, Denmark \\
$^{22}$ Dept. of Physics, TU Dortmund University, D-44221 Dortmund, Germany \\
% ...
$^{35}$ Dept. of Physics and Astronomy, University of Kansas, Lawrence, KS 66045, USA \\
% ...
$^{39}$ Dept. of Physics and Wisconsin IceCube Particle Astrophysics Center, University of Wisconsin{\textemdash}Madison, Madison, WI 53706, USA \\
% ...
$^{41}$ Department of Physics, Marquette University, Milwaukee, WI 53201, USA \\
% ...
$^{43}$ Bartol Research Institute and Dept. of Physics and Astronomy, University of Delaware, Newark, DE 19716, USA \\
% ...
$^{63}$ Deutsches Elektronen-Synchrotron DESY, Platanenallee 6, D-15738 Zeuthen, Germany \\
$^{\rm a}$ also at Institute of Physics, Sachivalaya Marg, Sainik School Post, Bhubaneswar 751005, India \\
% ...

\subsection*{Acknowledgments}

\noindent
The authors gratefully acknowledge the support from the following agencies and institutions:
USA {\textendash} U.S. National Science Foundation-Office of Polar Programs,
U.S. National Science Foundation-Physics Division,
U.S. National Science Foundation-EPSCoR,
U.S. National Science Foundation-Office of Advanced Cyberinfrastructure,
Wisconsin Alumni Research Foundation,
Center for High Throughput Computing (CHTC) at the University of Wisconsin{\textendash}Madison,
Open Science Grid (OSG),
Partnership to Advance Throughput Computing (PATh),
Advanced Cyberinfrastructure Coordination Ecosystem: Services {\&} Support (ACCESS),
Frontera and Ranch computing project at the Texas Advanced Computing Center,
U.S. Department of Energy-National Energy Research Scientific Computing Center,
Particle astrophysics research computing center at the University of Maryland,
Institute for Cyber-Enabled Research at Michigan State University,
Astroparticle physics computational facility at Marquette University,
NVIDIA Corporation,
and Google Cloud Platform;
Belgium {\textendash} Funds for Scientific Research (FRS-FNRS and FWO),
FWO Odysseus and Big Science programmes,
and Belgian Federal Science Policy Office (Belspo);
Germany {\textendash} Bundesministerium f{\"u}r Forschung, Technologie und Raumfahrt (BMFTR),
Deutsche Forschungsgemeinschaft (DFG),
Helmholtz Alliance for Astroparticle Physics (HAP),
Initiative and Networking Fund of the Helmholtz Association,
Deutsches Elektronen Synchrotron (DESY),
and High Performance Computing cluster of the RWTH Aachen;
Sweden {\textendash} Swedish Research Council,
Swedish Polar Research Secretariat,
Swedish National Infrastructure for Computing (SNIC),
and Knut and Alice Wallenberg Foundation;
European Union {\textendash} EGI Advanced Computing for research;
Australia {\textendash} Australian Research Council;
Canada {\textendash} Natural Sciences and Engineering Research Council of Canada,
Calcul Qu{\'e}bec, Compute Ontario, Canada Foundation for Innovation, WestGrid, and Digital Research Alliance of Canada;
Denmark {\textendash} Villum Fonden, Carlsberg Foundation, and European Commission;
New Zealand {\textendash} Marsden Fund;
Japan {\textendash} Japan Society for Promotion of Science (JSPS)
and Institute for Global Prominent Research (IGPR) of Chiba University;
Korea {\textendash} National Research Foundation of Korea (NRF);
Switzerland {\textendash} Swiss National Science Foundation (SNSF).

\end{document}